# A Reservoir Model of Explicit Human Intelligence

Eric C. Wong

Departments of Radiology and Psychiatry

University of California, San Diego

ecwong@ucsd.edu

**Abstract**

A fundamental feature of human intelligence is that we accumulate and transfer knowledge as a society and across generations. We describe here a network architecture for the human brain that may support this feature and suggest that two key innovations were the ability to consider an offline model of the world, and the use of language to record and communicate knowledge within this model. We propose that these two innovations, together with pre-existing mechanisms for associative learning, allowed us to develop a conceptually simple associative network that operates like a reservoir of attractors and can learn in a rapid, flexible, and robust manner. We hypothesize that explicit human intelligence is based primarily on this type of network, which works in conjunction with older and likely more complex deep networks that perform sensory, motor, and other implicit forms of processing.

**Introduction**

Modern human intelligence is qualitatively different from that of other animals in that we accumulate societal intelligence across generations in our writings, memory, and other media. For most modern human adults, the vast majority of their knowledge and understanding is not generated by individual experience with the natural environment, but learned from other humans, and this education process is now the focus of the first few decades of life.

This qualitative difference compared to lower animals suggests that human specific aspects of intelligence may not be implemented by a quantitative improvement of the networks that support lower-level intelligence, but instead by utilizing or exploiting different mechanisms or architectures in the brain to allow for the accumulation and utilization of collective societal intelligence.

The brains of animals include networks that process incoming sensory streams into abstractions. For example, the visual system transforms raw input through layers of increasing abstraction into representations of entities in the environment. There are also output networks that convert abstracted outputs such as 'walk' into parallel signals for motor coordination. In between these input and output systems lie networks that operate on abstractions themselves. In simple cases an input abstraction may be connected directly to an output abstraction, such as a connection between 'predator' and 'run', but in many cases, and especially for human cognition, processing among abstractions is more complex.

Human specific intelligence operates primarily on discrete definable abstractions, most of which are labeled using words, which facilitate both storage and communication of knowledge. The portion of human knowledge that is explainable and therefore readily accumulated is referred to as explicit knowledge [1] and is described by reportable relationships between abstractions. Such relationships are likely to be stored in a network of abstractions that excite one another as discrete reportable entities, as opposed to a network that performs processing through hidden layers or representations. While this is not necessarily the case, any non-trivial processing that occurs through hidden representations is not directly reportable and can only be taught through post-hoc examination of our processing.

It is well known that direct associations between nearly arbitrary abstractions are easily produced in the brain, as exemplified by classical conditioning [2], and it is straightforward to cast most of human specific knowledge in the form of associations - such as between a word and its definition, between a right triangle and the Pythagorean theorem, or between a person and a historical event. A network architecture that is composed of abstractions with direct, learnable connections is consistent with both the accumulable, reportable nature of explicit human intelligence and with known associative mechanisms.

Knowledge that is not completely reportable is referred to as implicit knowledge [1] and includes skills such as riding a bike which can only partly be taught and requires practice, as well as social intelligence such as reading the mood of others, and sensory processing such as visual object identification. We propose that implicit knowledge in humans is supported primarily through mechanisms inherited from lower animals that have evolved over millions of years, but that the explicit, human specific aspects of intelligence may be supported using simple associative mechanisms that build upon pre-existing mechanisms for learning associations, but leverage new functionalities, enabled by the development of imagination and subsequently language, to allow for rapid societal accumulation of human intelligence.

We outline here a simple working hypothesis for a network architecture that may support explicit human thought.

**A Reservoir of Attractors**

We hypothesize that human specific cognition is supported primarily by a conceptually simple associative network architecture that behaves in principle like a reservoir network [3], but one in which the reservoir itself learns in a flexible, robust, and potentially deep manner.

Reservoir computing refers to an architecture in which input is delivered to a large network with random recurrent connections, known as a reservoir, that performs random transformations on the input. This network is connected to an output layer, and during supervised training, only the connections between the reservoir and the output layer are modified. The concept is that the reservoir performs all possible relevant transformations on the input data, up to a single layer transform, and this final transform is learned during training. The training process can be viewed as an associative learning step between those elements of the reservoir that were active just prior to the presentation of the desired output, and the active elements of the output.

In the associative network that we hypothesize, we assume that abstractions such as objects, concepts, and events are each represented in the brain by high activity in a group of neurons that we will refer to as an attractor [4], also referred to in other work as ensembles or cell assemblies [5-7], or engrams [8, 9]. We have proposed elsewhere a mechanistic theory for how attractors may be both formed and associated with one another through generic spike timing dependent plasticity (STDP) mechanisms [10].

We assume that the elemental units of the reservoir are attractors rather than neurons, and that therefore the reservoir is a network of relationships between abstractions, which can in principle be transparent and reportable, rather than opaque relationships between hidden representations. We also assume that the collection of attractors is both the reservoir and the output. In the operation of the network, a small number of attractors is excited, either by sensory input or by other attractors in the case of internal thought, and a sequence of excitation ensues, limited in breadth by attention mechanisms, and resulting either in an output action or ongoing internal thought. In the reservoir the path from input to output can be as shallow or as deep as appropriate for any task.

During training an input is provided, exciting the corresponding attractors, and the desired output is subsequently specified by a teacher, which may be another human, text, a diagram, or another source of accumulated human knowledge. The input is processed through the intelligence that had previously been encoded into the connections of the reservoir, resulting in the excitation of attractors corresponding to the output of that pre-existing intelligence. When the attractors corresponding to the specified output are excited by the teacher, these are attached to the attractors excited by the input through associative learning using the same mechanism as in classical conditioning. This represents learning of the new information, as the same input can now be processed into the new desired output.

Unlike a conventional reservoir network, the output of the proposed network is the excitation of attractors within the reservoir. Therefore, new associations, representing new intelligence, are immediately available to participate in flexible future learning. While curricular learning is typically hierarchical in nature [11], the proposed learning method can make use of any attractors that are excited by the input to connect to the desired output attractors, whether the resulting pattern of connections produces a hierarchical structure or not.

When learning requires the construction of a new attractor, such as the learning of a new abstraction and a corresponding word, an engram containing references to the multisensory input during teaching is initially stored in the hippocampus using episodic memory mechanisms [12] and subsequently generates a corresponding new attractor in the reservoir, possibly by the mechanisms proposed in [10]. The new attractor would initially represent an episodic memory in the sense that it can excite all of the multisensory information that was encoded into the engram, but over time, if the most relevant associations between the new abstraction and related attractors are strengthened through co-excitation, then other episodic information such as the identity of the teacher or the time of day may be lost, leaving only connections representing semantic content. This scenario is consistent with previous suggestions that all memory is initially episodic, while only some becomes semanticized [13, 14].

Because the proposed architecture is a simple associative network of attractors, the primary structure of the network can be described by a graph or connectivity matrix. However, we imagine that the dynamics of the network are shaped by highly nonlinear forms of interaction within and between attractors, including modulation by neuromodulators, local feedback and homeostatic mechanisms via myriad neuronal and glial cell types, and attentional modulation. Through these interactions the edges of the graph would be described not by simple connection weights, but by more complex temporal and learning characteristics. We do not propose here a detailed model for these interactions, but as an example, a basic attention mechanism might be as simple as transient lateral inhibition by excited attractors, producing competition that limits highly excited attractors to a small number. With such a mechanism in place, the sequence of attractor excitations could be relatively confined to attractors that are most relevant to the present context, and plastic changes along one sequence could be isolated from other sequences or contexts. Without this kind of mechanism excitation would likely spread widely across the large reservoir and gravitate toward the eigenvectors of the global weight matrix with the highest eigenvalues, producing far less varied dynamics. We suggest that the sequence of attractor activations that exceeds a threshold set by attention mechanisms is the subject of conscious awareness and subject to episodic memory, and is the sequence which is known informally as the train of thought or internal monologue. It is also possible that working memory is implemented by the excitation of attractors into a state that provides more prolonged excitation.

The proposed network contrasts with deep artificial neural networks, where the nodes and connections are usually simple in behavior, representations are typically distributed codes rather than confined to discrete attractors, intermediate processing occurs through hidden layers or representations, and the opportunity for high levels of intelligence appears to depend on synaptic plasticity that is broadly coordinated across the network. We believe that the proposed structure is more consistent with communication and societal accumulation of intelligence as well as few-shot learning [15], and resistance to catastrophic forgetting [16]. Because the proposed network utilizes simple association, few-shot learning is relatively straightforward, as new associations would naturally build bridges between areas of the underlying graph that previously had low connectivity. For example, after learning that an animal is a mammal, the attractor representing this animal would then have close connections to attractors representing all properties that were previously associated with mammals and excite or partially excite those attractors upon future encounters with the animal. Similarly, all properties of the animal would be added to the list of properties associated with mammals. Catastrophic forgetting is likely to occur when learning of new functions requires widespread modification of synapses that are shared across different functions. If attention mechanisms restrict excitation to a narrow set of attractors that are relevant to the current task, as we suggest, then synaptic modification would be restricted to that set as well, largely avoiding perturbation of connections related to other functional domains.

**Implicit intelligence**

Examples of implicit (or non-declarative) intelligence might include riding a bicycle, or when an experienced chess player has an intuitive sense that one position is better than another but cannot produce an explicit explanation for why that is so. If knowledge is not explainable, or incompletely explainable, then at least some portion of the processing is hidden from the conscious awareness of the human. Within the model described here this can occur in at least two different ways. In the case of riding a bicycle, the output is the coordinated contraction of many muscles, with real-time feedback from multiple sensory systems. This coordination is learned mainly through practice, with success and failure providing feedback, consistent with reinforcement learning in deep networks which is not amenable to complete description. In a case such as intuitive chess knowledge, it may be that the underlying processing is again mediated by a deep network, but it is also possible that it is mediated by associations between discrete attractors in the reservoir network described above, representing moves or positions, and is only hidden from the player's awareness because it involves partial excitation of many attractors, but each below the level required for conscious attention. With practice, the associative pathways may be refined so that the most relevant attractors rise high enough in activity to become subjects of attention, and the knowledge thereby transitions from implicit to explicit.

While we propose that explicit human intelligence may be implemented largely by a reservoir of attractors, this network must be supported by implicit functionalities that are inherited from lower animals and likely more genetically pre-programmed through evolution. These include systems for sensory processing and motor coordination, as well as intuitive physics and intuitive psychology [17] systems that implicitly model physical and social aspects of the world.

**Imagination**

A fundamental function that is supported by associative learning is the conversion of observations into predictions. When events A and B occur in sequence, associative learning can construct a directional connection between the attractors representing A and B, such that in the future the excitation of attractor A excites or potentiates attractor B, directly implementing evidence-based prediction. This is useful because it allows animals to react quickly to anticipated future events. A new level of cognition is achieved when a prediction can be used not only to drive behavior in a feedforward manner, but to model the world farther into the future by allowing a prediction to serve as a hypothetical input. Once this critical ability is supported the brain can operate independently of the present environment and support processing within an offline world model consisting of hypothetical events, actions, and predictions. This also requires the brain to be able to switch reliably between two distinct modes: interacting with the environment through sensory input and motor output, and hypothetical processing within this offline model that allows for prediction from hypothetical actions without motor output.

The development of neural mechanisms to support imagination was a critical evolutionary step to enable human intelligence and required the development of robust mechanisms to allow for safe and robust switching between imaginary thought and externally directed processing. While these mechanisms are not understood in detail, there is now a large body of neuroimaging data showing that the human default mode network is specifically associated with internal thought [18], and therefore that the distinction between internal and externally directed thought has a neurobiological basis.

**Language**

The use of language is critical for both the communication and the societal accumulation of human intelligence. Words are not explicitly necessary for either communication or imagination, as information can be communicated through gestures, example behavior or art, and imagination can also be internally visualized objects or events. However, with words assigned as labels for anything we can think of, sequences of words can be used as a proxy for demonstration that is compact, teachable, and recordable, and also allows for hierarchical definition to produce higher levels of abstraction such as mathematical theorems. Words also allow one person to directly excite arbitrary sequences of attractors in the brain of another, in effect providing immediate access to elements of the world model of the listener. In the proposed reservoir the tens of thousands of words we learn would presumably serve as the backbone of the network and also as the addressing system for random access to the network.

The sequence of attractor excitations that represents internal thought is closely related to, but not identical to, the expression of a thought using language. Multiple attractors can be excited simultaneously, like chords in music, while language is necessarily linear. This linearization is a central function of our language system [19], and we must also learn in the other direction to parse language into the intended series of attractor excitations. While we suggest that the language system is critical to the expansion of human intelligence through its role in communication and accumulation of knowledge in the proposed reservoir network, we believe that the language system itself is more likely supported by deep networks that learn by reinforcement learning during the early years of language acquisition to perform these complex grammatical and syntactic transformations. In the initial phases of development and language acquisition, children learn grammar as well as skills such as what constitutes a statement or a question, when to start and stop talking, and how to translate between language and the internal language of thought. These skills are learned gradually over the preschool years through repetition and practice and are consistent with reinforcement-based learning on top of genetically determined circuitry. However, once these skills are learned children progress to a second phase in which a distillation of human knowledge is systematically constructed into the reservoir through teaching. The initially shallow associative network of attractors eventually becomes a deep reservoir network of attractors, but because it is an associative network of attractors, the knowledge that resides in the connections remains largely explainable.

**Discussion**

The central question addressed here is what brain mechanisms have developed to support the distributed human intelligence that we observe today. We argue here that the nature of the recent increase in human intelligence, characterized by rapid societal accumulation and dissemination of explainable intelligence suggests network mechanisms that are tailored to these characteristics. In particular, we suggest that the benefits of communicability require that the state space of the proposed reservoir network be restricted to the activity level of discrete reportable attractors, as opposed to the much larger state space but relative opacity of a less constrained network of similar size. This results in the nonintuitive idea that high level cognitive function in humans may be simpler in concept and more compact to describe than lower level functions such as visual processing and motor coordination.

We have outlined a network architecture that we believe is consistent with the above properties and propose that the additional enabling innovations in brain biology were mechanisms that support wakeful imagination and language. We suggest that these along with pre-existing associative learning mechanisms allow us to build and modify one another's reservoir networks, and thereby extend the scope of the entity developing human intelligence beyond the individual brain to the society.

One prediction of the proposed model is that within the reservoir there are no hidden layers or representations. This provides a potential point of comparison with experimental data, though this is complicated by the possibility that sub-threshold excitation of attractors is common, and that only a subset of the neurons that comprise an attractor may fire for any given excitation.

For a computational model of the proposed network, we suggest that there are too many unknown properties that arise from microstructure, attention, and feedback mechanisms to construct a computationally accurate model from the bottom up, and that a more useful approach would be to parameterize the dynamics and plasticity of the reservoir heuristically. If a parameterization is found that produces effective learning in simulation, then the resulting dynamics may serve as a potential target for more detailed neurobiological modelling.

In addition to the reservoir itself, a functional computational model would require the simulation of at least a minimal set of support networks. Ideally, sensory and motor systems would interact with the reservoir at the level of reportable abstractions, and a reward system would provide valence and modulate plasticity, these being largely equivalent to the intuitive physics and psychology engines described by Lake et al [17]. Finally, the language of thought, which in this model may be identical to the sequence of attractor excitations, would need to be mapped to conventional language for the purpose of communication.

**Acknowledgements**
We would like to thank Richard Buxton, Thomas Liu, Peter Bandettini, and Alan Simmons for many helpful discussions regarding the subject of this paper.